\documentclass[a4paper,onecolumn,11pt]{quantumarticle}
\pdfoutput=1
\usepackage[utf8]{inputenc}
\usepackage[english]{babel}
\usepackage[T1]{fontenc}
\usepackage{amsmath}
\usepackage{hyperref}

\usepackage{tikz}
\usepackage{lipsum}

\usepackage{amssymb}
\usepackage{epsf,float,epsfig}
\usepackage[numbers,sort&compress]{natbib}

\begin{document}

\title{Beyond Worst-Case Branching: Quantum Tree Search via Amplitude Amplification}

\author{Andreas Wichert}
\affiliation{Department of Computer Science and Engineering,  INESC-ID  \&  Instituto Superior T\'ecnico, University of Lisbon, {2740-122} Porto Salvo, Portugal}
\orcid{0000-0002-2179-4378}
\email{andreas.wichert@tecnico.ulisboa.pt}
\homepage{http://web.ist.utl.pt/andreas.wichert/}
\orcid{0000-0003-0290-4698}
%\thanks{You can use the \texttt{\textbackslash{}email}, \texttt{\textbackslash{}homepage}, and \texttt{\textbackslash{}thanks} commands to add additional information for the preceding \texttt{\textbackslash{}author}. If applicable, this can also be used to indicate that a work has previously been published in conference proceedings.}

\maketitle

\begin{abstract}
In this work, we investigate quantum tree search via amplitude amplification. Amplitude amplification generalizes Grover's algorithm by replacing the Hadamard initialization with an arbitrary unitary $A$, with Grover's algorithm recovered as the special case of uniform initialization. We demonstrate the construction of a dynamic search tree of depth $m$ with query complexity  $\sqrt{\left(b_{avg}\right)^m}$ where $b_{avg}$  denotes the average branching factor, improving upon the commonly assumed $\sqrt{\left(b_{max}\right)^m}$, where $b_{max}$ is the maximum branching factor. We further challenge the widespread assumption that amplitude amplification is inferior to quantum backtracking. In fact, quantum backtracking is unsuitable for problems that do not naturally admit a backtracking structure; in such cases, amplitude amplification yields improved query complexity.
We observe that amplitude amplification constructs the search tree dynamically, rendering its internal structure inaccessible, a constraint that applies equally to quantum backtracking. To address this, we propose sampling-based methods to estimate the tree structure, under the assumption that it approximates a normal distribution with increasing depth. Finally, we introduce a quantum greedy search based on a lookahead heuristic inspired by the classical cognitive architecture Soar, which models human-like problem-solving strategies.
 \end{abstract}

%Keywords:  Quantum Tree Search; Amplitude Amplification; Grover's Algorithm;  Non Constant Branching; Quantum Heuristics; 

\maketitle

\section{Introduction}

In this work, we study quantum tree search algorithms through the lens of amplitude amplification, with the aim of clarifying and sharpening complexity claims that have been imprecisely stated in the existing literature  \cite{Rennela2923}.  To ground our theoretical development concretely, we illustrate all key concepts and algorithms using the 8-puzzle, a canonical sliding tile problem whose search tree exhibits the kind of variable, state-dependent branching structure that makes the distinction between average and worst-case complexity practically meaningful.

A central contribution is a rigorous analysis of query complexity for dynamically constructed search trees. Specifically, we demonstrate that for a search tree of depth $m$, the query complexity is governed by  $\sqrt{\left(b_{avg}\right)^m}$, where $b_{avg}$ denotes the average branching factor of the tree of depth $m$, a bound that is strictly better than the commonly assumed worst-case complexity based on the maximum branching factor $b_{\max}$  \cite{tarrataca2010}. This distinction is non-trivial: conflating average- and worst-case branching factors leads to a systematic overestimation of the cost of amplitude-amplification-based search, and has contributed to misleading comparisons with alternative quantum approaches   \cite{Rennela2923}.

We address a recurring claim in the quantum search literature, namely that Grover-based algorithms are generically inferior to quantum backtracking \cite{Rennela2923}. We argue that this comparison is only valid when the problem structure naturally admits a backtracking formulation,  that is, when the search space can be organized as a tree with well-defined pruning conditions that quantum walk can exploit. For problems that do not naturally form a backtracking tree, this assumption breaks down, and amplitude amplification can achieve a strictly better query complexity than quantum backtracking.
A second structural observation concerns the dynamic nature of the search tree. The algorithm constructs the search tree. This dynamic construction means that the internal structure of the tree is not accessible as an explicit data structure at any point during the computation. Importantly, this is not a limitation peculiar to our approach: the same constraint applies to quantum backtracking, where the implicit tree explored by the quantum walk is similarly inaccessible. 
Given that the tree structure is not directly observable, we propose statistical methods to estimate it from the outside. These estimates can be used to set the number of amplitude amplification iterations.

Finally, we propose a quantum greedy search algorithm based on a lookahead heuristic. The design of this heuristic is motivated by the classical cognitive architecture Soar \cite{Laird87}, a well-studied model of human-like problem solving that combines goal-directed search. By incorporating a quantum lookahead step, evaluating multiple candidate branches in superposition before committing to a direction,  the algorithm emulates the deliberative, heuristic-guided reasoning that characterizes effective human search behavior. 
Together, these contributions provide a more complete and accurate picture of quantum tree search via amplitude amplification: correcting complexity claims, characterizing structural limitations, providing tools to work around those limitations, and extending the framework in a cognitively motivated direction. 

Throughout, the 8-puzzle serves as a concrete running example, grounding abstract complexity results in a tractable, well-understood domain.

\section{Tree Search and Grover's Algorithm}

\begin{figure}[htb]
%\vspace{15cm}
\begin{center}
\leavevmode
\epsfysize3.4cm
\epsffile{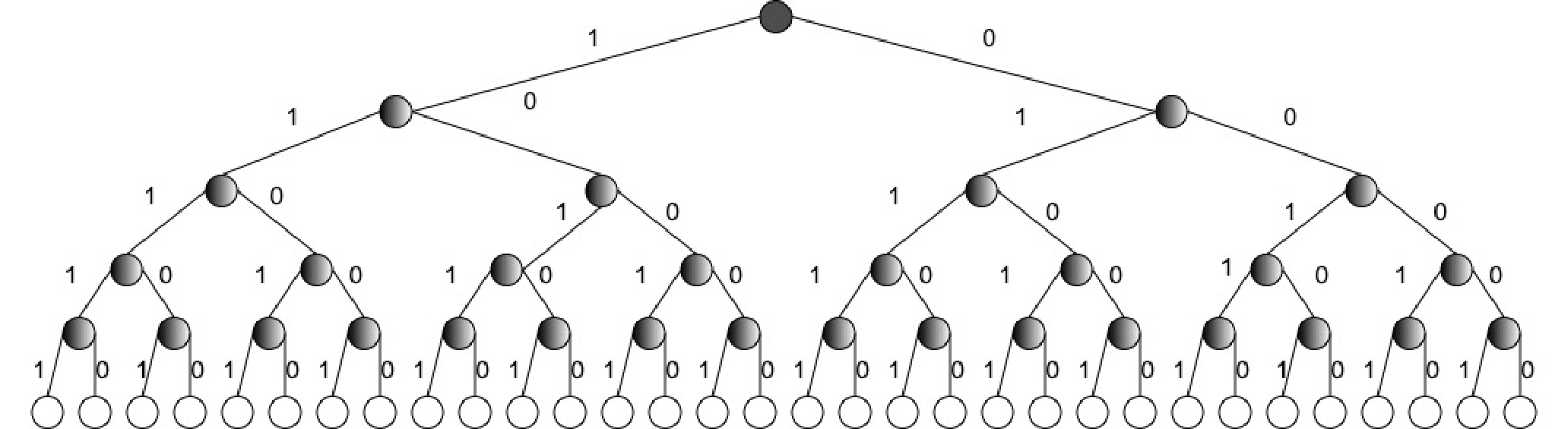}
\end{center}
\caption{Search tree for $b=2$ and depth  $m=5$ with $32= 2^5$ leaf nodes. Each unique path from the root to a leaf node is represented by one of the $32$ binary numbers, each consisting of five digits. Each binary number serves as a path descriptor, with each digit indicating whether to move left (1) or right (0).}
\label{Tree_2_3.eps}
\end{figure}

Nodes and edges in a search tree represent states and transitions between states. The initial state is the root, and from each state $v$, either $b=br(v)$ states can be reached or it is a leaf. From a leaf, no other state can be reached,  $br(v)=0$. We call $b=br(v)$ the branching factor of a node $v$, indicating the number of possible choices. A leaf can either be the desired of the computation or an impasse when there are no valid transitions to a succeeding state.
Each node $v$ representing a state except the root has a unique parent node. Every node and its parent are connected by an edge, and
each parent has $b=br(v)$ children.  
A path in a graph is an ordered sequence of nodes, where each pair of adjacent nodes in the sequence is connected by an edge from the root node to a leaf node. Each path represents a unique route from the root node to the leaf node.
The depth of a node $v$ is defined as the number of edges on the path from the root node to that node.
The depth of the search tree is indicated by $m$. It is defined as the number of edges that connect the root node to the deepest leaf node.  
When the branching factor is constant, and each leaf node has the same depth, the tree exhibits exponential growth in relation to its depth $m$.
For instance, when the branching factor $b$ is constant at $2$ and the depth of the tree $m$ is $5$, the resulting search tree is depicted in Figure \ref{Tree_2_3.eps}. 
This tree has $32= 2^5$ leaf nodes. Each unique path from the root to a leaf node is represented by one of the 32 binary numbers, each consisting of five digits. Each binary number serves as a path descriptor, with each digit indicating whether to move left (1) or right (0). 
For a constant branching factor $b>2$, each question has $b$ possible answers. The $m$ answers can be represented by a base-$b$ number with $m$ digits.
In a quantum computation, we can simultaneously represent all possible path descriptors. There’s one path descriptor for each leaf off the tree.
Using Grover’s algorithm, we search through all possible paths and verify whether each path leads to the desired state that is marked by an oracle. This type of procedure is known as a quantum tree search \cite{tarrataca2010}. For $n=b^m$ possible paths, the costs are (approximately) $\sqrt{n}=b^\frac{m}{2}$ corresponding to the reduced branching factor $b_q=\sqrt{b}=b^\frac{m}{2}$ by Grover's search\footnote {In our analysis we will not use the $O$ notation
and for simplification we will assume the approximate complexity.}. If the depth $m$ off the search tree is not known, iterative deepening can be applied. In iterative deepening search, the search limit is progressively increased from one to two to three to four, and so on, until a desired is discovered. For each limit, a search is conducted from the root to the maximum depth of the search tree. If the search proves unsuccessful, a new search is initiated with a deeper limit.
During iterative deepening search, states are generated multiple times \cite{korf1985}, \cite{russell2010}.
The time complexity of iterative deepening search is comparable to that of a search to the maximum depth \cite{korf1985}. A quantum iterative deepening search is equivalent to iterative deepening search \cite{tarrataca2012b}, \cite{Tarrataca2013}.
For each limit $max$, a quantum tree search is performed from the root, where $max$ represents the maximum depth of the search tree. The potential solutions are determined through measurement. 

\section{Generalized Tree Search}

\begin{figure}[htb]
%\vspace{15cm}
\begin{center}
\leavevmode
\epsfysize2cm
\epsffile{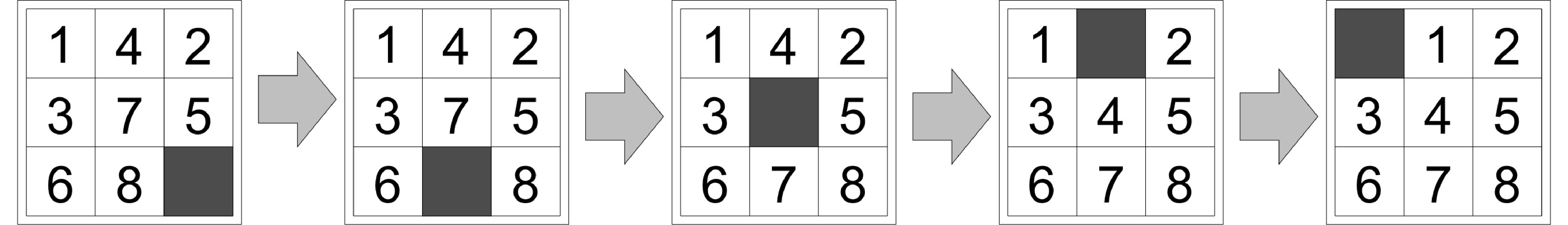}
\caption{The first pattern (left) represents the initial state and the last (right) the desired state. The series of moves describe a possible solution to the problem. }
\label{Search8.eps}
\end{center}
\end{figure}

We introduce the fundamental concept of generalized tree search  tree search with variable branching factor by the 8-puzzle game.
The game consists of eight numbered movable tiles arranged within a $3 \times 3$ frame. One of the cells is unoccupied, enabling tiles to be relocated to generate various patterns. The objective is to identify a sequence of tile movements that relocate the tiles into the empty space, transforming the board from its initial state to a desired state (as illustrated in Figure \ref{Search8.eps}).
\begin{figure}[htb]
%\vspace{15cm}
\begin{center}
\leavevmode
\epsfysize6cm
\epsffile{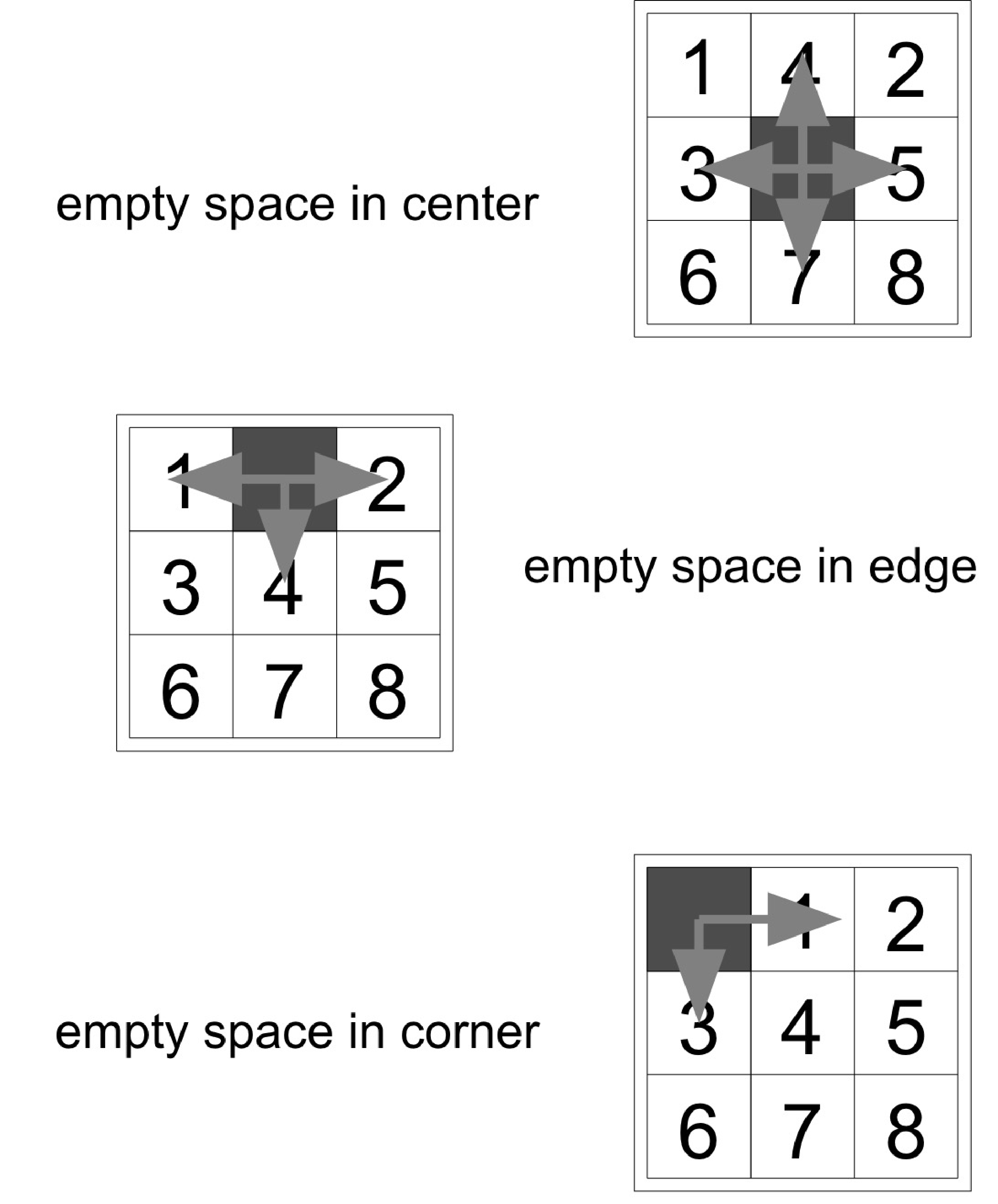}
\caption{With the empty space in the center, the empty tile can move right, left, up, or down. For the empty spaces on the edge there three possible movements and for the empty spaces in the corner  two possible move }
\label{Moves}
\end{center}
\end{figure}
With the empty space in the center, the empty tile can move right, left, up, or down. For the empty spaces on the edge there three possible movements and for the empty spaces in the corner  two possible move, (see Figure \ref{Moves}). Therefore, $b_{max}=4$, however, the average branching factor is $b_{avg}=2.6667$. This is because there are nine possible positions,
\begin{equation}
b_{avg}=\frac{1 \cdot 4 + 4 \cdot 3 + 4\cdot 2}{9}=2.6667.
\end{equation}
and there is one possible position of empty space in the center, and four possible positions of empty space in edge or corner.
Generally we describe that the branching factor depends on categories with
\[
b_{\text{avg}}=\sum_{i} p_i\, b_i,
\]
where  $b_i$ = branching factor for category $C_i$ and $p_i $= probability/fraction of being in category $C_i$. In 8-puzzle example there three categories $C_i$,
\begin{itemize}
\item $C_1$ empty space in the center, $b_1=4$, $p_1=1/9$.
\item $C_2$ empty spaces on the edge, $b_2=3$, $p_1=4/9$.
 \item $C_3$ empty spaces in the corner, $b_3=2$, $p_1=4/9$.
 \end{itemize}
\[
b_{avg}=\frac{1}{9} \cdot (4)+\frac{4}{9}  \cdot (3)+\frac{4}{9}  \cdot (2)=2.6667.
\]
with $b_1= b_{max}=4$.

\subsection{Generalized Quantum Tree Search}

The generalize quantum tree search can be characterized through two distinct approaches which build the search tree dynamically
\begin{itemize}
\item Dynamic Pumping: Generates an uniform initial distribution, enabling the application of Grover’s amplification.
\item Dynamic Superposition: This technique generates a non-uniform distribution, necessitating amplitude amplification as suggested by Gilles Brassard et al. \cite{brassard2000}. 
\end{itemize}
Both approaches dynamically construct the quantum search tree, rendering it impossible to access its internal structure.
They provide a dual description of the same problem. The dynamical pumping approach is simpler to implement, as standard Grover’s amplification can be utilized.
The dynamical superposition approach is more intuitive to comprehend, as the generated superpositions correspond to the branching factors of the nodes.

\subsection{Dynamic Pumping}

In the absence of a constant branching factor at each level of the tree, with uniform leaf depth, the maximal branching factor $b_{max}$ must be employed for the quantum tree search as proposed by  \cite{tarrataca2010}. 
For a node $v$, if the value of $b(v)$ is less than the maximum allowable value $b_{max}$, only $b(v)$ determines the moves. 
The remaining $b_{max} - b(v)$ local paths are not utilized. We enhance the path descriptors by repeating existing moves within the unused local paths. This repetition corresponds to the dynamic construction of the search tree. We refer to this method as dynamic pumping \cite{Wichert2022b}.
Given that we are utilizing qubits, we can approximate the maximum number of moves as follows
\[\hat b_{max}=2^{\lceil \log_2(b_{max}) \rceil}. \]
and
\[\hat b_i=2^{\lceil \log_2(b_i) \rceil}. \]
For a node $v$ in category $C_i$ with $\hat b_i < \hat b_{max}$, only $b_i$ moves are required. The remaining moves, $\hat b_{max} - \hat b_i$, would correspond to unused moves.
We augment the moves by repeating the same moves with
\begin{equation}
rep_i=\left \lfloor \frac{\hat b_{max}}{\hat b_i}  \right \rfloor  = \hat b_{max} \div  \hat b_i
\end{equation}
and  we are guaranteed to have $rep_i$ repetition of paths containing a solution. 
In the case
\begin{equation}
\left \lfloor \frac{\hat b_{max}}{\hat b_i}  \right \rfloor \neq \frac{\hat b_{max}}{\hat b_i},  
\end{equation}
\begin{equation}
reminder_i=\hat b_{max} - rep_i \cdot \hat b_i =\hat b_{max} \mod \hat b_i.
\end{equation}
we are left additionally with $reminder_i$ unused moves. 
If we additionally include non-solution moves, we can no longer guarantee that the augmented paths contain a solution. There are two approaches to handle the case
 where $reminder_i \neq 0$.

\subsubsection{Deterministic Padding} 

\begin{figure}[htb!]
\begin{center}
\leavevmode
\epsfysize3cm
\epsffile{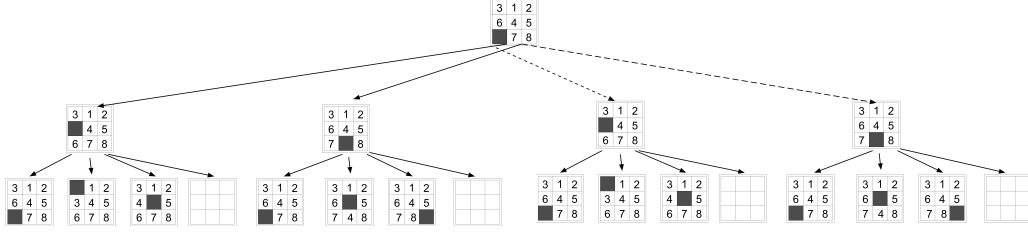}
\caption{Example of a search tree for 8-puzzle using dynamic pumping and the deterministic padding. From the initial state there are two possible moves, we repeat the two moves indicated by dashed line from the root. In the deterministic approach the reminder resulting from three possible moves is filled by an empty representation of the state that is not used (garbage).}
\label{PumpDet.eps}
\end{center}
\end{figure}

In this scenario, we designate the corresponding path with $reminder_i \neq 0$ as a garbage state (a non existing state in the problem domain) that does not contribute to the search process and is subsequently repeated to the search, see Figure \ref{PumpDet.eps}.
The  average branching factor is denoted as
\[
\hat b_{\text{avg}}=\sum_{i} p_i\, \hat  b_i.
\]
By discarding the $reminder_i$   paths, the approach becomes deterministic, since we are guaranteed exactly  $rep_i$  repetitions of solution containing paths.
In the 8-puzzle example the category $C_1$ and $C_2$ have the same branching value due to qubit representation.
\begin{itemize}
\item $C_1$ empty space in the center, $\hat b_1=4$, $p_1=1/9$.
\item $C_2$ empty spaces on the edge,  $\hat b_2=4$, $p_1=4/9$ , since $4=2^{\lceil \log_2(3) \rceil} =2^{2}. $
 \item $C_3$ empty spaces in the corner, $\hat b_3=2$, $p_1=4/9$.
 \end{itemize}
with 
\[
\hat b_{avg}=\frac{1}{9} \cdot (4)+\frac{4}{9} \cdot (4)+\frac{4}{9}  \cdot (2) =3.1111
\]
The primary drawback of this approach is that the remaining unused paths contribute to the retrieval cost

\subsubsection{Probabilistic Padding}

\begin{figure}[htb!]
%\vspace{15cm}
\begin{center}
\leavevmode
\epsfysize3cm
\epsffile{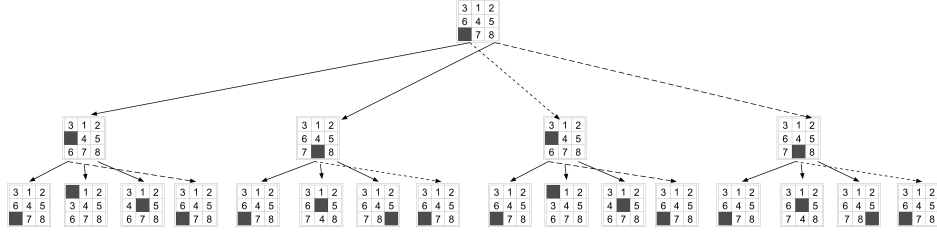}
\caption{Example of a search tree for 8-puzzle using dynamic pumping and the probabilistic padding. From the initial state there are two possible moves, we repeat the two moves indicated by dashed line from the root. In the probabilistic approach the reminder resulting from three possible moves is filled by a repetition of a state, in our case the leftmost move indicated by dashed line from the nodes of depth one.}
\label{PumpProb.eps}
\end{center}
\end{figure}

The selected remainder moves can further contribute to the repetition of solution-containing paths, better approximating the true average branching factor.
\begin{equation}
b_{\text{avg}}=\sum_{i} p_i\, (\hat  b_i - reminder_i)  \approx \sum_{i} p_i\, b_i.
\end{equation}
Since reminder values are employed to generate some additional paths to leaves, this approach  is probabilistic in relation to retrieval costs, see Figure \ref{PumpProb.eps}. 
There can be multiple repetitions of paths that may or may not contain a solution, since the probability of selecting a rule on the path to the solution for the $reminder_i$ moves is not guaranteed. The probabilistic padding approach seeks to mitigate the limitations of qubit representation by ensuring the desired outcome, provided that the representation constraints are not binding. In the 8-puzzle example we assume $b_{avg}=2.6667$.

\subsubsection{Iterations}

For $k$ solutions, the probability of measuring a state that represents one solution of $k$ solutions is directly proportional to the number of iterations of the Grover’s operator, denoted by $r$. The probability of observing a solution should be as close to $1$ as possible, while the number of iterations, $r$, should be minimized. After approximately $r$ iterations, the probability of measuring a solution approaches $1$~\cite{hirvensalo2004,nielsen2000} with
\begin{equation}
r=\left\lfloor \frac{\pi}{4} \cdot \sqrt{\frac{2^m}{k}}\right\rfloor.
 \end{equation} 
The number of iterations $r$ is the largest integer not exceeding the computed value.

For a tree of depth $m$, there are on average $\hat \kappa_{avg}$ copies of the subtree, or equivalently, on average $\hat \kappa_{avg}$ repletions represent the same leaves
\begin{equation}
\hat \kappa_{avg}=\left(\frac{\hat b_{max}} {\hat b_{avg}}\right)^m .
\end{equation}
Since we will have different path descriptors leading to the same solution, we require on average a lower number of Grover’s iterations for the correct solution and 
we can state that
\begin{equation}
\hat r _{avg} \approx \sqrt{\frac{\left(\hat b_{max}\right)^m}{\hat \kappa_{avg}}} =   \sqrt{\left( \frac{\hat b_{max}^m}{\left(\frac{\hat b_{max}} {\hat  b_{avg}}\right)}\right)^m} = \sqrt{\left(\hat  b_{avg} \right)^m} 
\end{equation}
or, depending on the $reminder_i$ approach
\[
r_{avg} \approx \sqrt{\left( b_{avg} \right)^m} .
\]
The validity of this solution is contingent upon the average repetition on the path that is $\hat \kappa_{avg}$ times. 
The saving compare to $O(\sqrt{\left( b_{max} \right)^m})$ is considerable since using the ``O'' notation we can show that for
\[
 b_{avg} < b_{max},
\]
\begin{equation}
O \left(\sqrt{\left( b_{avg} \right)^m} \right) \neq O\left(\sqrt{\left( b_{max} \right)^m} \right).
\end{equation}
since we assume in other case that there exists a constant $c$, that for certain value $m_0 > 0$ 
\[
0 \leq \left( b_{max} \right)^m \leq c \cdot \left( b_{avg} \right)^m~~  \forall  m > m_0.
\]
However such a constant does not exist becasuse
\[
\left( \frac{b_{max}}{b_{avg}} \right) ^m \leq c
\]
follows the contradiction
\[
m \cdot \log \left( \frac{b_{max}}{b_{avg}} \right)\leq \log (c),
\]
\[
m \leq \log \left( c -  \frac{b_{max}}{b_{avg}} \right).
\]

\subsection{Dynamic Superposition}

\begin{figure}[htb]
%\vspace{15cm}
\begin{center}
\leavevmode
\epsfysize3cm
\epsffile{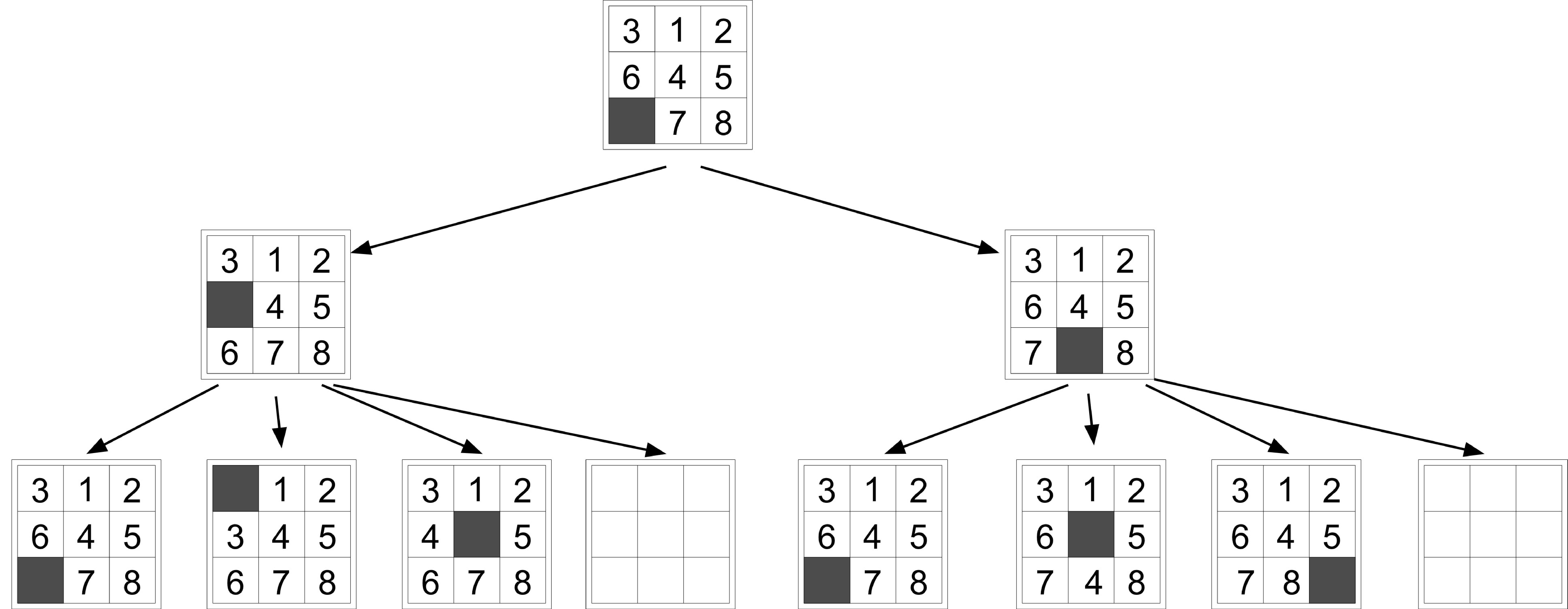}
\caption{Example of a search tree for 8-puzzle using dynamic superposition and the deterministic padding. In the deterministic approach the reminder resulting from three possible moves is filled by an empty representation of the state that is not used (garbage).}
\label{ChosenDet.eps}
\end{center}
\end{figure}

\begin{figure}[htb]
%\vspace{15cm}
\begin{center}
\leavevmode
\epsfysize3cm
\epsffile{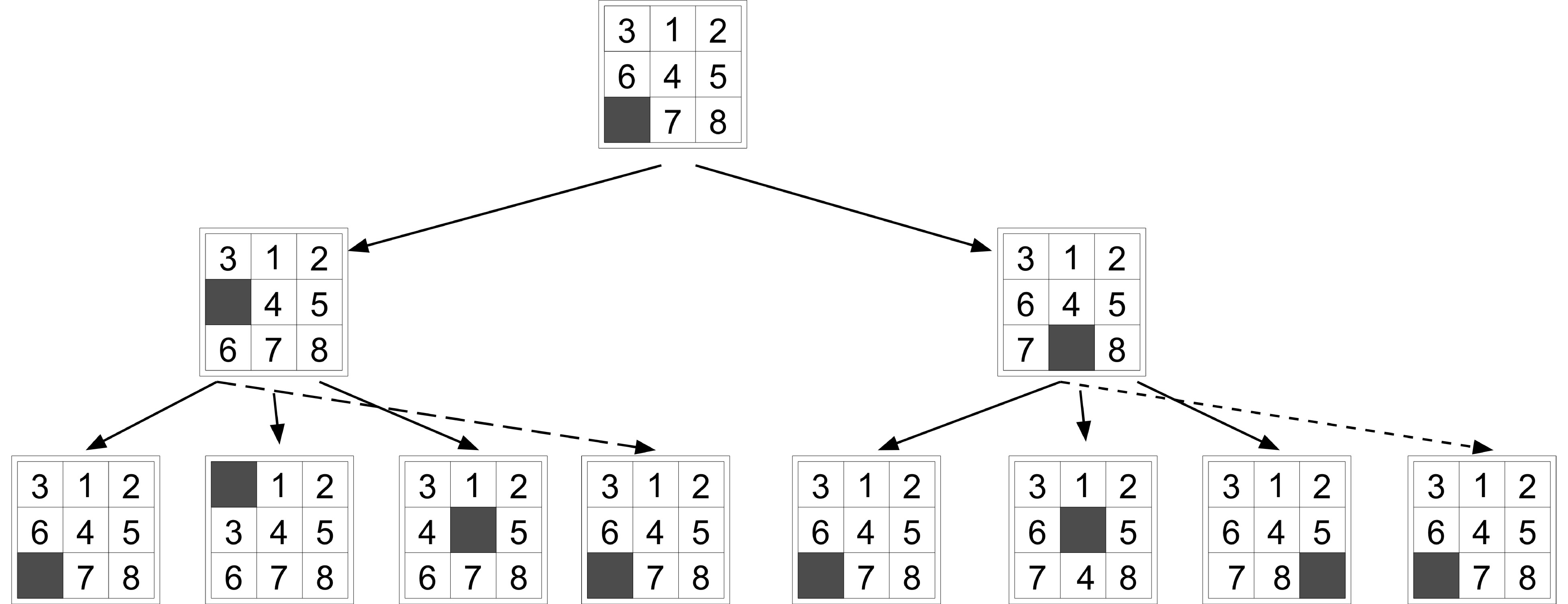}
\caption{Example of a search tree for 8-puzzle using dynamic superposition and the deterministic padding. In the probabilistic approach the reminder resulting from three possible moves is filled by a repetition of a state, in our case the leftmost move indicated by dashed line from the nodes of depth one.
}
\label{ChosenProb.eps}
\end{center}
\end{figure}

An equivalent approach employing variable superposition was proposed in \cite{Sequeira2021}.
For a node $v$ in category $C_i$, a superposition is generated by
\begin{equation}
 |\psi_v \rangle = \frac{1}{\sqrt{b_i}} \cdot \sum_{v \rightarrow x} |x\rangle  =  \frac{1}{\sqrt{b_i}} \cdot  \sum_{i=1}^{b_i} |x_i\rangle 
\end{equation}
since we are using qubits with $\hat b_i=2^{\lceil \log_2(b_i) \rceil}$ with $\hat b_i \leq \hat b_{max}$ 
\begin{equation}
 |\psi_v \rangle =  \frac{1}{\sqrt{\hat b_i}} \cdot  \sum_{i=1}^{\hat b_i} |x_i\rangle = H^{\otimes \left(\frac{\hat b_i}{2} \right)}.
 \end{equation}
 with  $H^0=I$ being the identity operator. In the case where
 \begin{equation}
\hat b_{i} -b_i \neq  0
\end{equation}
 we can choose between the two approaches to address the case of reminder as before,
\[
reminder_i=\hat b_{max} - rep_i \cdot \hat b_i =\hat b_{max} \mod \hat b_i.
\]
either the deterministic or probabilistic padding, (see Figure \ref{ChosenDet.eps} and \ref{ChosenProb.eps}).
We generate the superposition by controlled Hadamard gates $cH$ by $ \hat b_i $ with $Pred(v)$ determining $| \hat b_i \rangle$ for the node $v$.
\begin{equation}
 | v \rangle | 0 \rangle =  | v \rangle Pred(v) | 0 \rangle = | v \rangle | \hat b_i \rangle =cH^{\otimes \left(\frac{\hat b_i}{2} \right)} | v \rangle | \hat b_i \rangle 
\end{equation}
which leads to entanglement of $|v\rangle|\hat{b}_i\rangle$.
We disentangle $|v\rangle|\hat{b}_i\rangle$ by recomputing with $Pred(v)^{-1}$ 
\begin{equation}
cH^{\otimes \left(\frac{\hat b_i}{2} \right)}   | v \rangle | \hat b_i \rangle =  |\psi_v \rangle | \hat b_i \rangle= |\psi_v \rangle Pred(v)^{-1}  | \hat b_i \rangle =  |\psi_v \rangle | 0 \rangle. 
\end{equation}
The resulting distribution of amplitudes is no longer uniform, as illustrated in the example of a search tree and the corresponding quantum circuit presented in Appendix \ref{ap:Variable Superposition}.
Assuming we have $\omega$ leaves with different depths, 
 \[ \mathcal L = \{\ell_1,\ell_2,\dots,\ell_j, \cdots, \ell_\omega\}. \]
A path $P_j$ from the root of the tree $v_1$ to a leaf $\ell_j$ is indicated as
  \[
  P_j = (v_1, v_{j,2}, \dots, v_{j,t}, \dots, \ell_j)
  \]
with $v_{j,t}$ is the index\footnote {$v_1 = v_{j,1}$ for all paths.}  chosen at depth $t$ and with  $d_j $ the depth of path $P_j$.
Each node that is not a leaf has a branching factor $\hat{b}(v_{j,t})$,  the amplitude of each leaf $\ell_j$ is given by
 \begin{equation}
 a(\ell_j)=\prod_{t=1}^{d_j} \frac{1}{\sqrt{\hat{b}(v_{j,t})}}
 \end{equation} 
 and is related to the branching factor and the depth. Assuming that the depth is  $m$ for all paths
  \begin{equation}
 a(\ell_j)= \prod_{t=1}^{m} \frac{1}{\sqrt{\hat{b}(v_{j,t})}}.
 \end{equation} 
 and we can define the path costs as
   \begin{equation}
 \hat \Omega_j  = \prod_{t=1}^{m}  \hat{b}(v_{j,t})= a(\ell_j)^{-2}.
 \end{equation} 
 Since the distribution $ |\psi\rangle$ generated by our algorithm $A$ that generated the quantum tree by dynamic superposition
\begin{equation}
A \cdot | 0^{\otimes m} \rangle = |\psi\rangle
  \end{equation} 
is not uniform 
\[
A \neq H^{\otimes m}
\]
we define the  diffuser as 
 \begin{equation}
2 \cdot |\psi \rangle\langle  \psi| - I_m  = A \cdot (2 \cdot |0^{\otimes m} - I_m) \cdot A^\dagger.
 \end{equation}
Amplitude amplification is performed with
 \begin{equation}
 |\psi\rangle =  a(\ell_{sol}) \cdot  | solution \rangle + ( 1- a(\ell_{sol}) )  \cdot | non~solution \rangle, 
  \end{equation}
after approximately $r$ iterations, the probability of measuring a solution approaches $1$ for
\begin{equation}
r=\left\lfloor \frac{\pi}{4} \cdot    \frac{1}{a(\ell_{sol}) } \right\rfloor.
 \end{equation} 
 The quantum amplitude amplification and estimation was proposed by Gilles Brassard  et al. \cite{brassard2000}.
 
 \subsection{Duality of Both Descriptions}
 
 Proof:  assume the solution is on path $P_j$ with path cost $\hat \Omega_j$
\[
\hat \Omega_j= \prod_{t=1}^{m}  \hat{b}(v_{j,t}),
\]
 so we need $\sqrt{\hat \Omega_j}$  iterations by variable superposition method.
 For dynamic pumping with $\hat b_{max}$ we get
 \begin{equation}
\hat \Gamma_j= a(\ell_j)^{-2}=\prod_{t=1}^{m}  \frac{\hat b_{max}}{ \hat{b}(v_{j,t})} \cdot  \hat{b}(v_{j,t})=\left( \hat b_{max} \right)^m
\end{equation}
 with
 \begin{equation}
 \hat \kappa_j=\prod_{t=1}^{m}  \frac{\hat b_{max}}{ \hat{b}(v_{j,t})} =\frac{\left(\hat b_{max}\right)^m}{\prod_{t=1}^{m}  \hat{b}(v_{j,t})}
\end{equation}
and 
\begin{equation}
\hat \Omega_j=\frac{\hat \Gamma_j}{\hat \kappa_j}=\frac{\left(\hat b_{max} \right)^m} {\frac{\left(\hat b_{max}\right)^m}{\prod_{t=1}^{m}  \hat{b}(v_{j,t})}} =\prod_{t=1}^{m}  \hat{b}(v_{j,t}).
\end{equation}

 As observed, the number of  iterations is inversely proportional to the square root of the probability value of the marked solution of the leaf of the tree. Both approaches exhibit isomorphic properties. In quantum dynamic pumping, the distribution is uniform, and lower branching is represented by the repetition of the same path. Conversely, in dynamic superposition, the distribution is non-uniform, and lower branching is represented by higher amplitudes. 
 
\subsection{Reconstruction of the Solution}

During the reconstruction of the solution, we must determine the branching factor.
After measurement, we obtain the binary path descriptor, which represents the path from the initial state to the desired state without specifying the chosen branching state. To reconstruct the path descriptor with the correct branching factor, we commence from the initial state and progressively ascertain the branching factor using the $Pred(v)$ operator.

\section{Sampling the Hidden Structure of the Search Tree}

Given that the quantum search tree is dynamically constructed, direct access to its structure is not feasible.
Consequently, we will explore the efficient estimation of the most probable path $P_s$ in which a solution with the amplitude $a(\ell_{sol})$ could be present.
We attempt to deduce the structure using some statistical tools based on prior information about the problem domain.
The  $\hat \kappa_{avg}$  repetitions need not be realized in practice, since they represent an average over all possible outcomes and  $\hat r_{avg}$ is not required to be a precise estimate. The actual number of $r$ iterations falls within the interval:
\begin{equation}
r \in \left[\sqrt{\left(\hat b_{min}\right)^m} , \sqrt{\left(\hat  b_{max}\right)^m} \right].
\end{equation}
The likelihood of encountering distinct paths varies significantly, as paths do not share the same probability, and not all values within the preceding interval are equally likely. The most probable outcome corresponds to the most frequent amplitude value among the leaf nodes, and we assume that with high probability the solution lies on the path with the most common amplitude value, which in turn determines the number of iterations. The distribution of path frequencies is taken to approach a normal distribution as $m$ grows large.

For $C$ categories of branching factors, the distribution can be described by a multinomial distribution, which generalizes the binomial distribution. The probability mass function is given by:
\[
\frac{n!}{x_1! \cdot x_2!\cdots x_C!}  \cdot   p_1^{x_1} \cdot  p_2^{x_2}\cdots p_k^{x_C}
\]
with  $x_i$ = number of outcomes in category $i$, $n = x_1 + x_2 + \cdots + x_C$, and $p_i$ = probability of category $C_i$.
Such a distribution can be readily estimated without simulation, as precise path descriptors are not required.
The approach simplifies further, since the binomial distribution is a special case of the multinomial. By grouping categories, a multinomial distribution can be collapsed into a binomial one with two categories.
The 8-puzzle can be directly described by a binomial distribution, as illustrated in Appendix \ref{ap: Binomial Distribution}, since each edge gives rise to exactly two distinct paths, each composed of a sequence of two moves.

\begin{figure}[htb]
%\vspace{15cm}
\begin{center}
\leavevmode
\epsfysize6cm
\epsffile{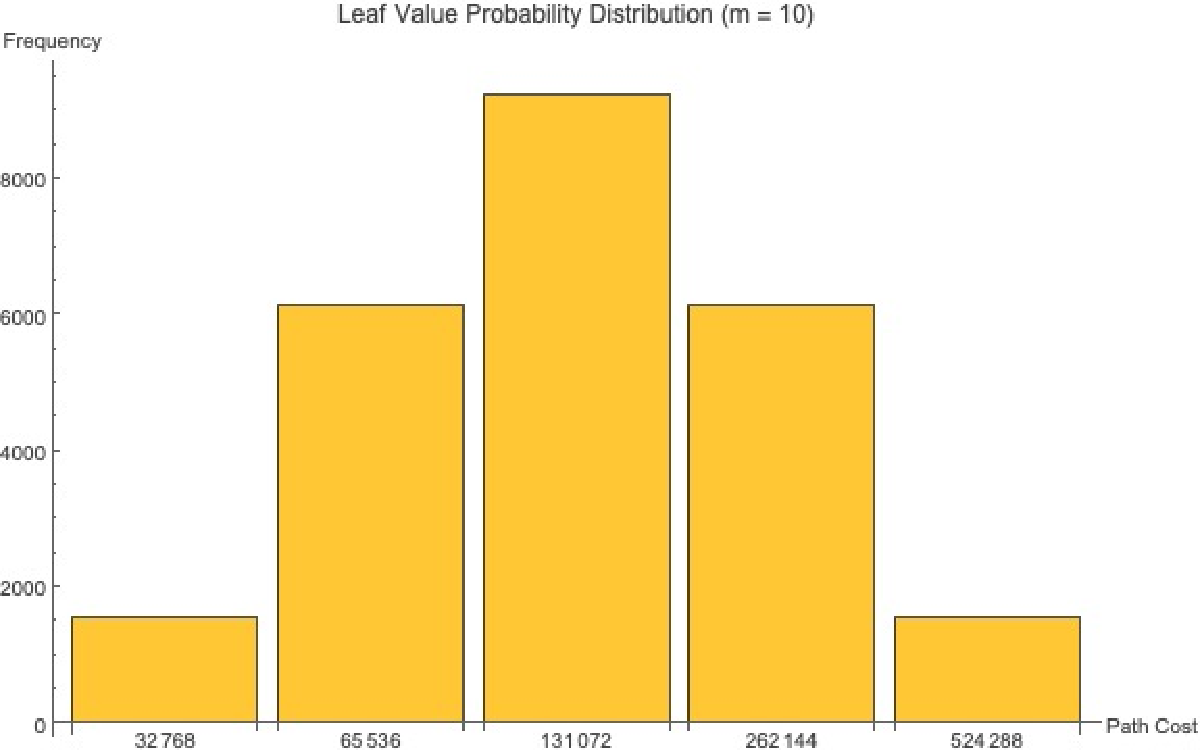}
\end{center}
\caption{Binomial distribution of the path cost frequency $\hat \Omega$  at depth  $m=10$ under the deterministic padding approach. The distribution for the probabilistic padding approach shares the same shape and frequency values, but is shifted toward lower path costs $\Omega$. }
\label{Binominal_10.eps}
\end{figure}

Figure \ref{Binominal_10.eps} depicts the binomial distribution for the 8-puzzle task at depth $m=10$. From the sampling  we can estimate the most frequent paths costs $\hat \Omega$ and the  correct value for $\hat r=\sqrt{\hat \Omega}$.

\subsection{Impact of Unique Solutions}

\begin{figure}[htb]
%\vspace{15cm}
\begin{center}
\leavevmode
\epsfysize6cm
\epsffile{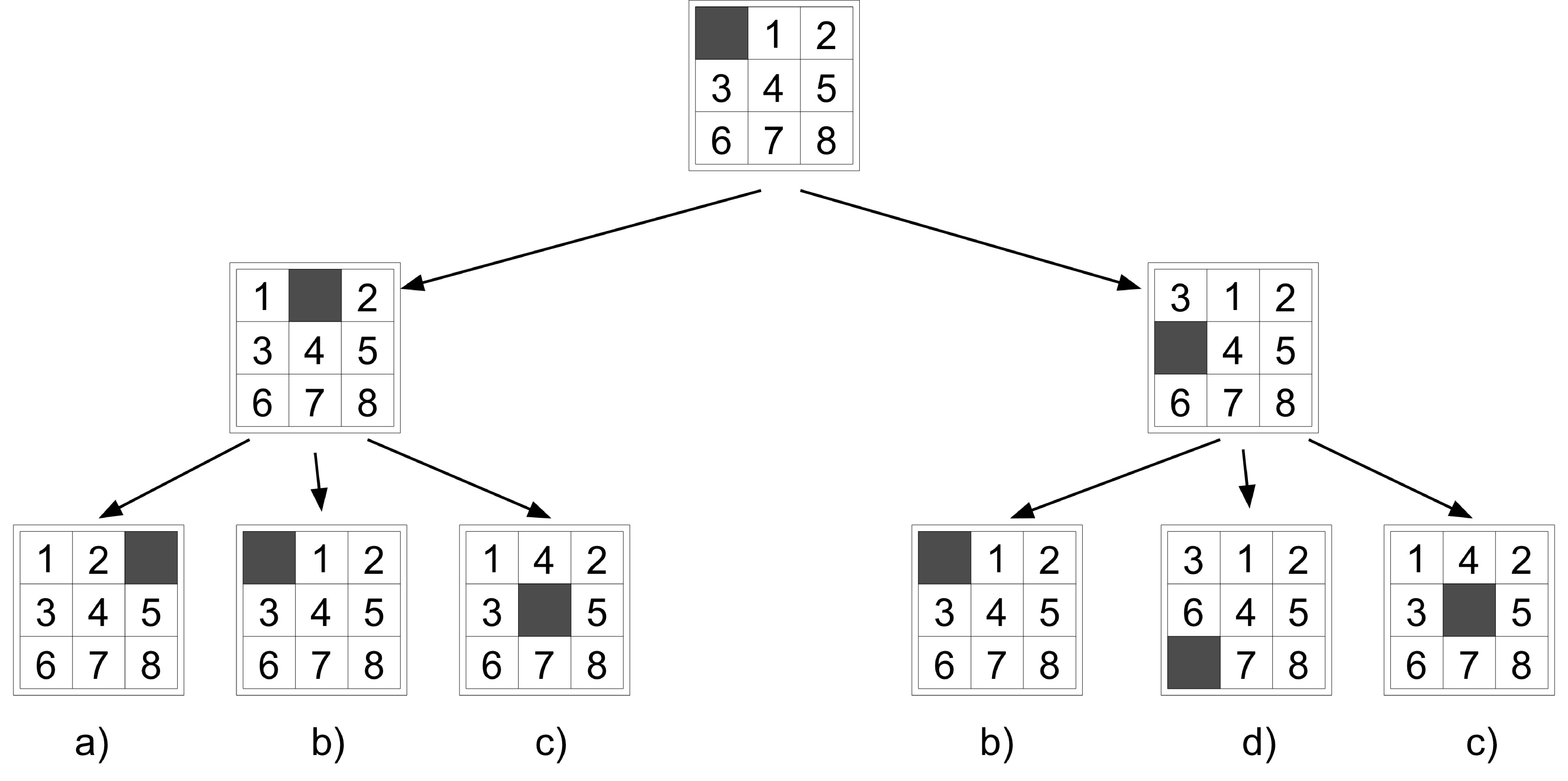}
\end{center}
\caption{Depth $m=2$ search tree for 8-puzzle. Since the problem is reversible, the reversible computation represents a solution path from the desired state (root) to different initial states (leaves), State a) and c) are represented once, while states b) and c) are represented twice.}
\label{Search-8puzzle.eps}
\end{figure}

A solution can be present multiple times since a state can be described by different paths. In Figure \ref{Search-8puzzle.eps}, we illustrate a simple search tree for an 8-puzzle of depth $m=2$. Four distinct states are generated. State a) and c) are represented once, while states b) and c) are represented twice.

\begin{figure}[htb]
%\vspace{15cm}
\begin{center}
\leavevmode
\epsfysize4cm
\epsffile{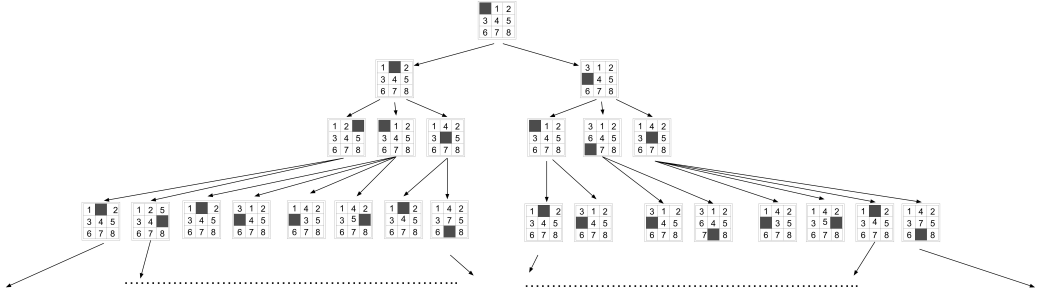}
\caption{For 8-puzzle we use Depth-First Search (DFS)  to explore all possible paths from the desired state to the leaves up to the defined depth $m$. Since the problem is reversible, the reversible computation represents a solution path from the desired state (root) to different initial states (leaves).}
\label{Search-8puzzle_deep.eps}
\end{center}
\end{figure}

To unify the path cost representation of state $X$  with path costs $\hat \Omega(X)$ and $\hat \kappa(X)$ repetitions,
we adopt a global notation $\hat \kappa_X$ and deduce
 \begin{equation}
\hat \Omega_X=\frac{\left( \hat b_{max} \right)^m}{\hat \kappa_X}.
 \end{equation}
 In the example of Figure \ref{Search-8puzzle.eps}  the cost is determined using a deterministic approach with with $\hat b_3=2$ and $\hat b_2=4$.
\[
\Omega(a)=8,~~\hat \kappa(a)=1~~\rightarrow~~\hat \kappa_a=\frac{4^2}{2 \cdot 4} \cdot 1=2,~~\hat \Omega_a=8,
\]
\[
\Omega(b)=8,~~\hat \kappa(b)=2~~\rightarrow~~\hat \kappa_b=\frac{4^2}{2 \cdot 4} \cdot 2=4,~~\hat \Omega_b=4,
\]
\[
\Omega(c)=8,~~\hat \kappa(c)=2~~\rightarrow~~\hat \kappa_c=\frac{4^2}{2 \cdot 4} \cdot 2=4,~~\hat \Omega_c=4,
\]
\[
\Omega(d)=8,~~\hat \kappa(d)=1~~\rightarrow~~ \hat \kappa_d=\frac{4^2}{2 \cdot 4} \cdot 1=2,~~\hat \Omega_d=8.
\]
A solution represented by state a) or b) requires one iteration, whereas states c) and d) require two iterations.
A general formulation is provided in the following: A state $X$ can have different paths indexed by $i$ with different costs $\hat \Omega(X_i)$ that can be repeated $\hat \kappa(X_i)$ times. The general formula used is given by 
 \begin{equation}
\hat \kappa_X=\sum_i \left( \frac{\left(\hat b_{max}\right)^m} { \hat \Omega(X_i) }  \cdot  \hat \kappa(X_i) \right).
 \label{eq:glkappa}
 \end{equation}
To estimate the impact of unique solutions, we conducted computer simulations.
For the 8-puzzle, we employed depth-first search (DFS) to explore all possible paths from the desired state to the leaves up to the specified depth $m$. Since the problem is reversible, the reversible computation represents a solution path from the desired stat to a different initial states, see Figure \ref{Search-8puzzle_deep.eps}. 

\begin{figure}[htb]
%\vspace{15cm}
\begin{center}
\leavevmode
\epsfysize9cm
\epsffile{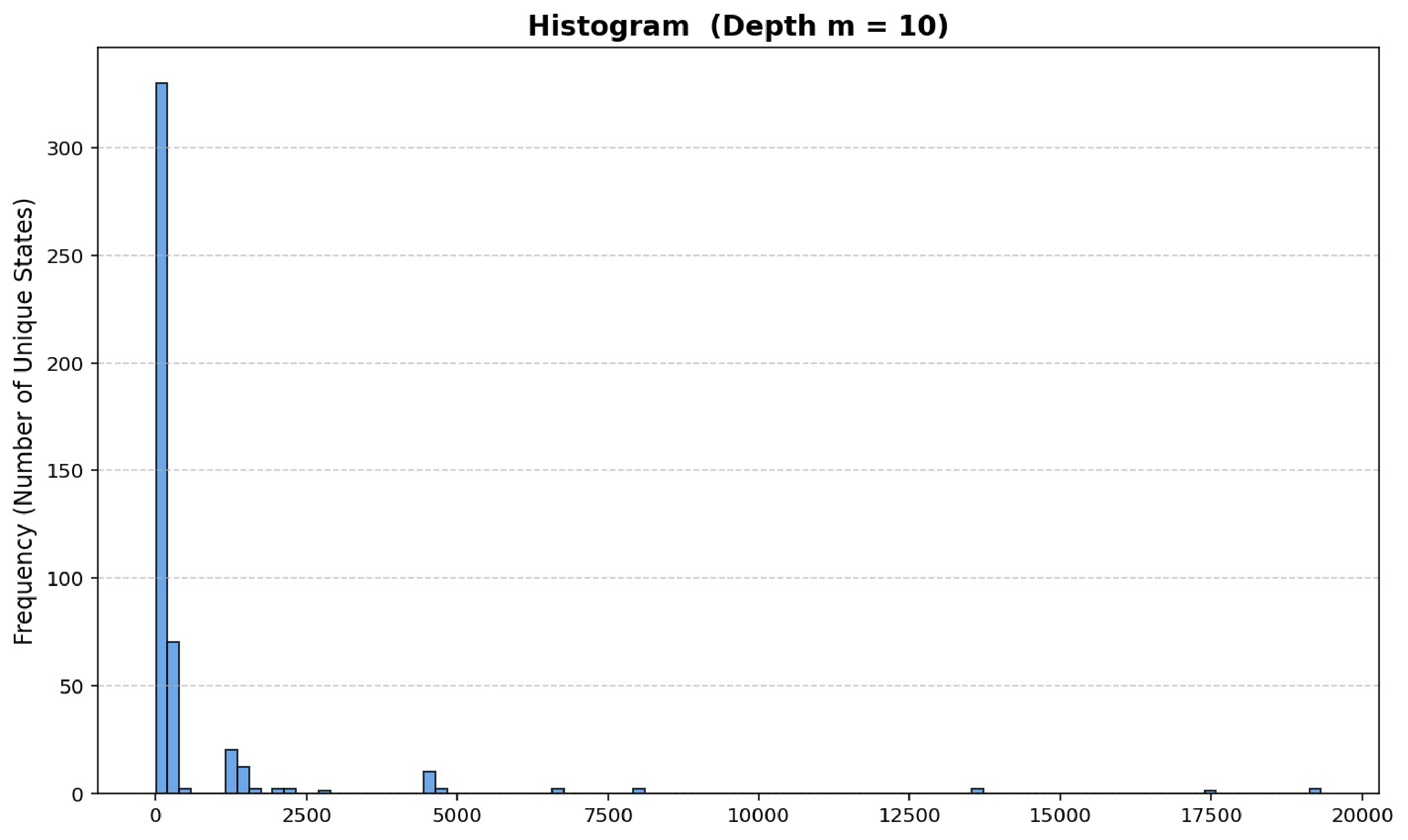}
\end{center}
\caption{Unique solutions for depth $m=10$ search tree for 8-puzzle. Employing Equation \ref{eq:glkappa}, we ascertain $\hat \kappa_X$ denoted on the x-axis for each distinct state and illustrate its frequency of occurrence. Notably, the most frequent value are $\hat \kappa_{X1}=8$ and $\hat \kappa_{X2}=16$. On the right side, we observe the proportion of $\hat \kappa$ values, which exhibits an upward trend, as numerous states emerged frequent with varying paths.}
\label{Depth_m_10_B.eps}
\end{figure}

Subsequently, using Equation \ref{eq:glkappa}, we determined $\hat \kappa_X$ for each unique state and indicated its frequency of occurrence. Refer to Figure \ref{Depth_m_10_B.eps} for a visual representation.
Table \ref{Binominal_10_DFS_Det} reports the values of  $\hat \Omega$  estimated by a binomial distribution, alongside the unique solutions found by DFS.
\begin{table} [h!]
\caption{For an 8-puzzle search tree of depth  $m=10$,  $\hat \Omega$  is estimated by a binomial distribution  with their corresponding appearance probabilities $Pr$. 
The table also reports the unique solutions $\hat \Omega$ found by depth-first search (DFS), together with their corresponding appearance probabilities $Pr$ (``-'' denotes no entry; ``$\cdots$'' additional data not shown). }
\begin{center}
\begin{tabular} { |c | c || c | c | }\hline
 \multicolumn{2}{|c||} {Binomial}   &  \multicolumn{2}{|c|}  {Unique Solutions}  \\ \hline
 $\hat \Omega$  &   $Pr$   &  $\hat \Omega$  &   $Pr$ \\ \hline \hline
   - & -  & $\cdots$ & $\cdots$ \\ \hline
 - & -   & 9362 & 0.0439\\ \hline
- & -   & 8738 & 0.05195\\ \hline
32768 &  0.0625  & 5041  & 0.0649\\ \hline
\textbf{65536}  & 0.25 &  \textbf{65536}  &  0.2035 \\ \hline
 \textbf{131072} & 0.375 & \textbf{131072}  & 0.3896 \\ \hline
 262144 & 0.25 & - & -  \\ \hline
 524288 &  0.0625 & - & - \\ \hline 
\end{tabular}
\end{center}
\label{Binominal_10_DFS_Det}
\end{table}
The probabilistic approach yields lower path cost $\Omega$  values. Additional results for the 8-puzzle task are provided in Appendix \ref{ap: DFS}.
To estimate a possible value of  $\hat r$  relative to  $\hat \Omega$  as indicated by the binomial distribution, we first take the most probable estimated value, then select the lesser of the two most probable values. Note that corresponding $\hat \Omega_{avg}=84945$ and that $ \Omega_{max}=1048576$ is  roughly two orders of magnitude larger.

\subsection{Impact of Backtracking}

If we reach an impasse without the subsequent state, we have two choices: propagate the corresponding state or propagate a garbage state. This process is commonly referred to as backtracking. Propagating the state to the leaf allows us to mark it as a potential solution, but without knowing its amplitude value. Backtracking performed in this na\"ive manner does not enable us to reduce the overall search costs, as the costs are related to the branching values on the path. For more advanced backtracking methods, please refer to the section  \ref{seq:Lookahead Backtracking}. 

\section{Quantum Greedy Search by Lookahead Heuristic}

The concept of cost savings through heuristics and amplitude amplification is not viable due to the inherent preparation cost associated with the initial distribution, as detailed in Appendix \ref{ap: NEA}.
Consequently, we propose an alternative approach that draws inspiration from the methodology employed in symbolic artificial intelligence based on Soar production system theory \cite{Laird87}.
Soar is a cognitive architecture designed to build symbolic AI systems that emulate human problem-solving. In Soar, the methodology is based on tree search, where all children generated from a node are examined before a child is selected as the next state. In our context, all children of state $v$ are generated through reversible computation without placing the generated states $\tilde x_i$  into superposition. The heuristic function $h(\tilde x_i)$ estimates the cheapest cost from node  to the goal state. 
 
We demonstrate two heuristic functions $h(\tilde x_i)$ applied to the 8-puzzle. Two widely used heuristics for this task are the number of misplaced tiles and the city-block distance \cite{Nilsson82,Pearl84,Luger98}. The first counts the number of tiles not in their goal position, but fails to account for how far each tile must travel. The city-block distance, also known as the Manhattan distance, instead sums the distances by which each tile is displaced, counting one unit for each square a tile must move to reach its goal position. The Manhattan distance is generally superior to the number of misplaced tiles as a heuristic. 

\begin{figure}[htb]
%\vspace{15cm}
\begin{center}
\leavevmode
\epsfysize8cm
\epsffile{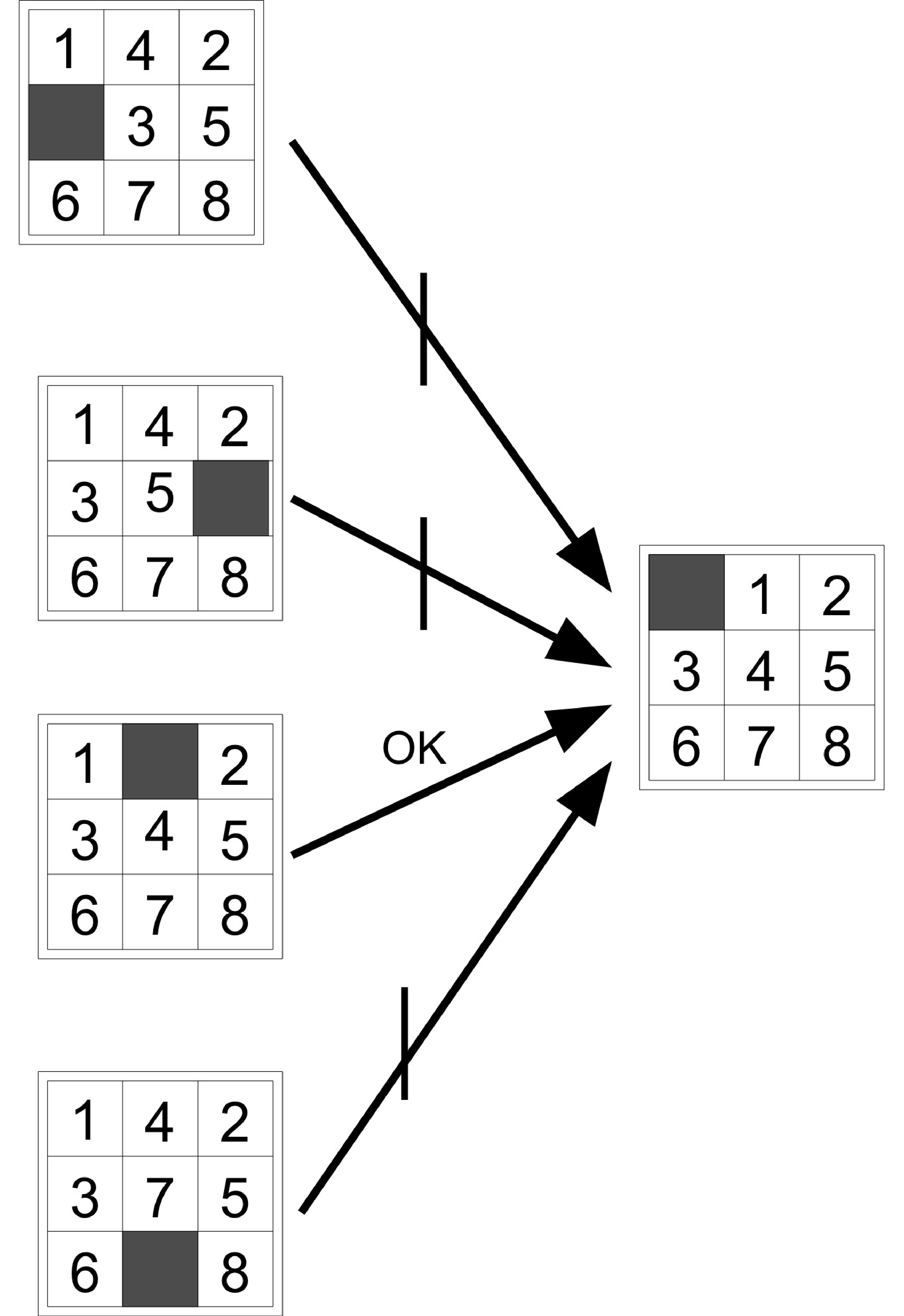}
\caption{All children of state $v$ are generated through reversible computation, without placing the generated states $\tilde x_i$  into superposition. The heuristic function $h(\tilde x_i)$ 
estimates the cost from each child node to the goal state. 
 Of the $\hat b_i$  children of  $v$, exactly one is selected for expansion, yielding an effective branching factor of 
 of $b=1$, In ourexample, four successor states are generated, and the heuristic identifies the one closest to the desired goal state marked “OK” in the figure.}
\label{Heuristic.eps}
\end{center}
\end{figure}

Of the $\hat b_i$  children of node $v$
\begin{equation}
 v \rightarrow  \sum_{i=1}^{\hat b_i} \tilde  x_i 
 \end{equation}
one child is selected by a heuristic function, which evaluates the corresponding states  $\tilde x_i$, 
and determines the one closest to the desired goal state via
\begin{equation}
\tilde x_l=\min_{1 \leq i \leq \hat b_i}(h(\tilde x_i))
\end{equation}
and  $\tilde x_i$ is  executed, 
\begin{equation}
 v \rightarrow  \tilde x_l
 \end{equation}
resulting in a branching factor of $b=1$, see Figure \ref{Heuristic.eps}.

For a tree of depth $m$, applying the corresponding heuristic on the $\alpha$ highest levels of the depth results in the costs of a  search tree of depth $m-\alpha$.
\begin{equation}
r_{avg} \approx  a(\ell_{avg})^ {-1}  \approx  \sqrt{\left(\hat b_{avg} \right)^{m-\alpha}}.
\end{equation} 
since
\begin{equation}
 a(\ell_j)= \prod_{t=1}^{m} \frac{1}{\sqrt{\hat{b}(v_{j,t})}}= \prod_{t=1}^{m-\alpha} \frac{1}{\sqrt{\hat{b}(v_{j,t})}}  \cdot \prod_{t=m-\alpha + 1}^{m} 1 = \prod_{t=1}^{m-\alpha} \frac{1}{\sqrt{\hat{b}(v_{j,t})}}.
 \end{equation} 
 Reaching a solution is not guaranteed, as a heuristic function may fail to identify a state that leads to the goal, potentially at substantial cost. To reconstruct the path to the solution, the same heuristic function must be applied uniformly at each level. The quality of the heuristic function significantly influences the search outcome, either yielding substantial savings or failing to find a solution, depending on the value of $\alpha$. The likelihood of failure increases with larger values of  $\alpha$  and a poorly chosen heuristic. For smaller values of $\alpha$, the states are closer to the goal, reducing the probability of failure.

\subsection{Lookahead Backtracking} \label{seq:Lookahead Backtracking} 

In the context of lookahead backtracking, we generate all possible children from a state $v$ through reversible computation without placing the generated states $\tilde x_i$ into superposition.
If a child or some children correspond to an impasse, we redefine the branching factor and execute the non-impasse children. This lookahead backtracking procedure enables us to reduce the overall search costs, as the costs are related to the redefined branching values on the path to the leaves.

\section{Discussion}

Quantum backtracking \cite{Montanaro2018} as well as the tree search via amplitude amplification as described in this paper  build dynamical the search tree which is unknown to the observer.
Both approaches require a prior knowledge to estimate its structure to query for the solution.
 In quantum backtracking, the cost of a quantum query is proportional to the number of nodes $T$ in the search tree and its depth $m$, resulting in an asymptotic complexity of $O(\sqrt{T} \cdot m^{3/2} \cdot \log m)$.
For relatively uniform search trees with an average branching factor $b_{avg}$, the number of nodes $T$ can be approximated as:
\[
T \approx \sum_{i=0}^{m} \left(b_{avg}\right)^i \approx \frac{\left(b_{avg}\right)^{m+1} - 1}{b_{avg} - 1}.
\]
This approximation yields a query complexity of approximately times
\[
\sqrt{\frac{\frac{\left(b_{avg}\right)^{m+1} - 1}{b_{avg} - 1}}{\left(b_{avg}\right)^m}} \cdot m^{3/2} \cdot \log m \approx \frac{b_{avg}}{b_{avg} - 1} \cdot m^{3/2} \cdot \log m
\]
than that of dynamic pumping, which relies on the simple Grover’s amplification. For $b_{avg}=2.6667$ and $m=10$, this would result in approximately $117$ times slower performance.

Furthermore, the amplitude amplification approach directly yields a path descriptor, whereas in quantum backtracking the path must be reconstructed by repeatedly applying the algorithm to subtrees to determine the correct branch at each level until the solution is found.

The choice between quantum backtracking and amplitude amplification depends on the problem structure.
For combinatorial problems with a natural backtracking tree structure \cite{Montanaro2018,Martiel2019,Rennela2923,Seidel2024} such as SAT, graph coloring, Sudoku, cryptanalysis, and branch-and-bound optimization,  quantum backtracking is preferable, since $T$, $m$ and the predicate cost can be directly modeled from the problem structure.

For problems that do not naturally admit a backtracking tree structure, such as the 8-puzzle, symbolic planning, and production systems, amplitude amplification is the more appropriate choice \cite{Wichert2022b,Wichert2024}.

\section{Conclusion}

The central message of this work is that no single quantum tree search algorithm dominates universally. The appropriate choice depends critically on the structure of the problem at hand. 

For problems whose search space admits clean decomposition into independent subtrees with effective pruning such as constraint satisfaction problems, quantum backtracking offers a well-motivated and theoretically grounded advantage, as the quantum walk can exploit the pruning structure directly \cite{Rennela2923}. However, this theoretical advantage must be qualified by a practical consideration that is too often set aside in the literature: quantum backtracking requires a number of qubits that scales with the depth and branching structure of the tree, and on current and near-term quantum hardware, the number of available, reliable qubits remains severely limited. When the problem is large enough that its backtracking tree demands more qubits than the hardware can support with acceptable fidelity, the theoretical quantum advantage evaporates entirely and a classical computer will in practice be faster. This is not a minor caveat. It implies that for many practically relevant problem sizes, quantum backtracking does not yet constitute a genuine alternative to classical search, regardless of its asymptotic behavior. The claimed advantage of quantum backtracking over amplitude amplification must therefore be understood as an asymptotic statement about a regime that present hardware cannot reach, rather than a practically actionable conclusion.

For problems whose structure is better characterized by a variable but bounded average branching factor, such as the 8-puzzle, amplitude amplification with complexity  $\sqrt{\left(b_{avg}\right)^m}$, provides a tighter and more practically relevant bound than backtracking-based analyses suggest. 
The 8-puzzle is a representative example, but the same analysis applies naturally to a broad class of state-space search problems in symbolic AI \cite{Nilsson82}, \cite{Winston92}, \cite{Luger98}.

Beyond the algorithmic contributions, a secondary but important contribution of this work is the introduction of new descriptive methods and analytical tools for reasoning about quantum search trees whose internal structure is not directly observable. 

When strong heuristic guidance is available, the quantum greedy lookahead variant can further reduce the effective search cost.\\
~\\
All examples can be freely downloaded at\\
 https://github.com/andrzejwichert/Tree-Search-via-AA.
\section{Acknowledgements}

 Work supported by national funds through Fundação para a Ciência e a Tecnologia, I.P. (FCT) under projects UID/50021/2025 (DOI: https://doi.org/10.54499/UID/50021/2025) and UID/PRR/50021/2025 (DOI: https://doi.org/10.54499/UID/PRR/50021/2025)
 
\section{Compliance with Ethical Standards}

The funders had no role in study design, data collection and analysis, decision to publish, or preparation of the manuscript. The authors declare no conflicts of interest. This article does not contain any studies with human participants or animals performed by any of the authors.

\appendix{}

\section{Appendix}

\subsection{Variable Superposition} \label{ap:Variable Superposition}

\begin{figure}[htb]
%\vspace{15cm}
\begin{center}
\leavevmode
\epsfysize3.4cm
\epsffile{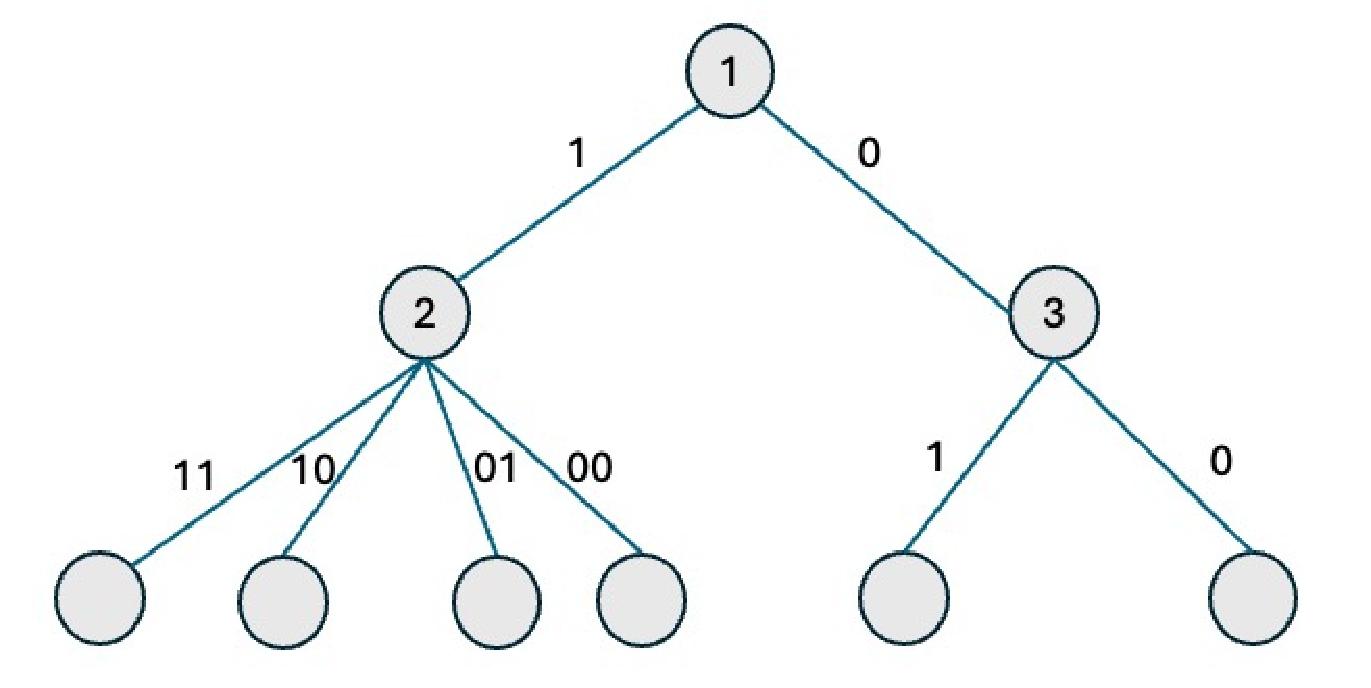}
\end{center}
\caption{A search tree of depth $m=2$ with $ \hat b_1=2$,  $\hat b_2=4$,  $\hat b_3=2$.}
\label{tree_m2.eps}
\end{figure}

\begin{figure} [htb]
%\vspace{15cm}
\leavevmode
%\epsfxsize\textwidth
%\epsfysize5cm
%\begin{center}
\parbox[b]{8cm}{ (a) \epsfxsize7cm\epsffile{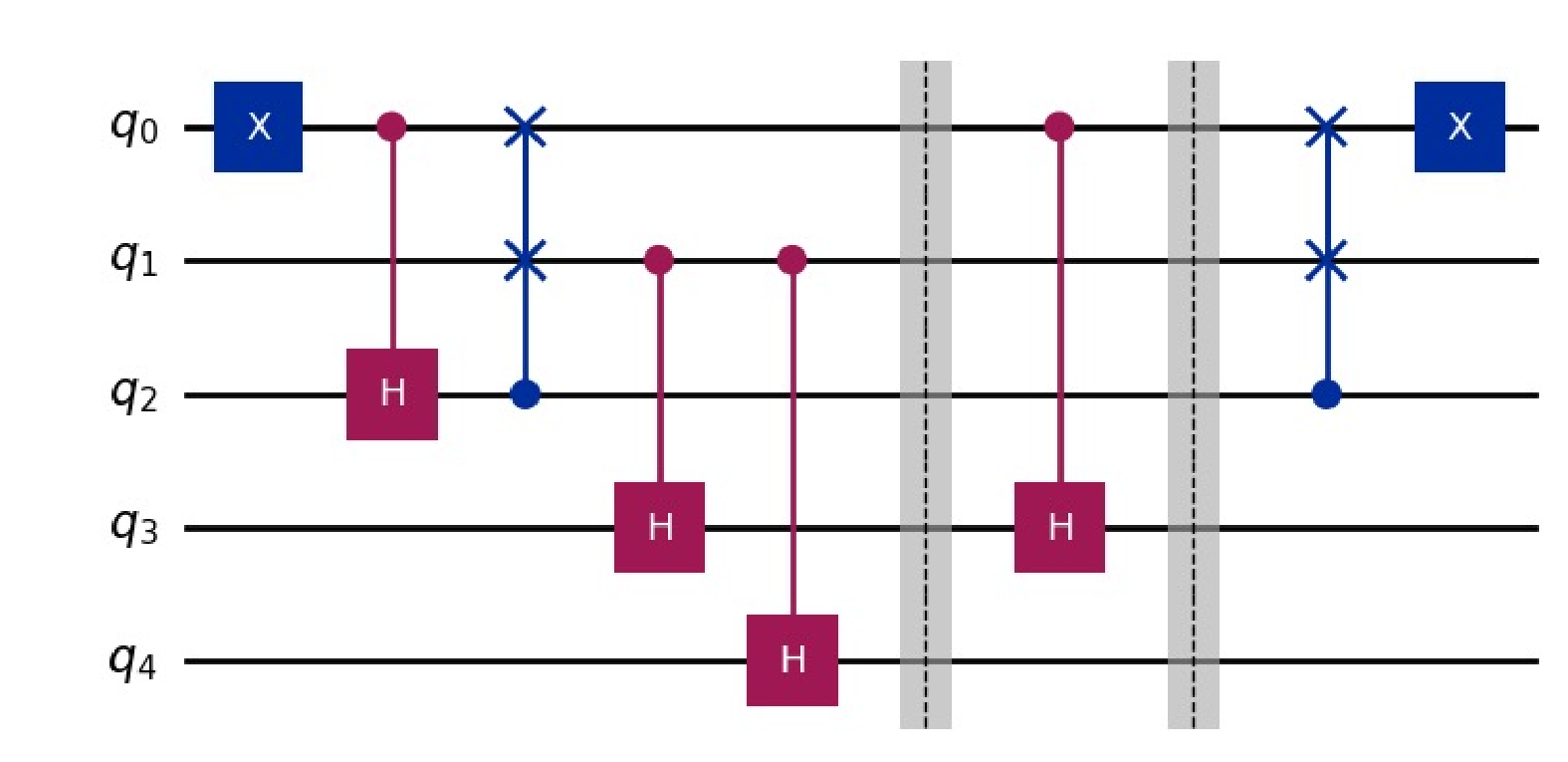}}
\parbox[b]{6cm}{ (b) \epsfxsize6cm\epsffile{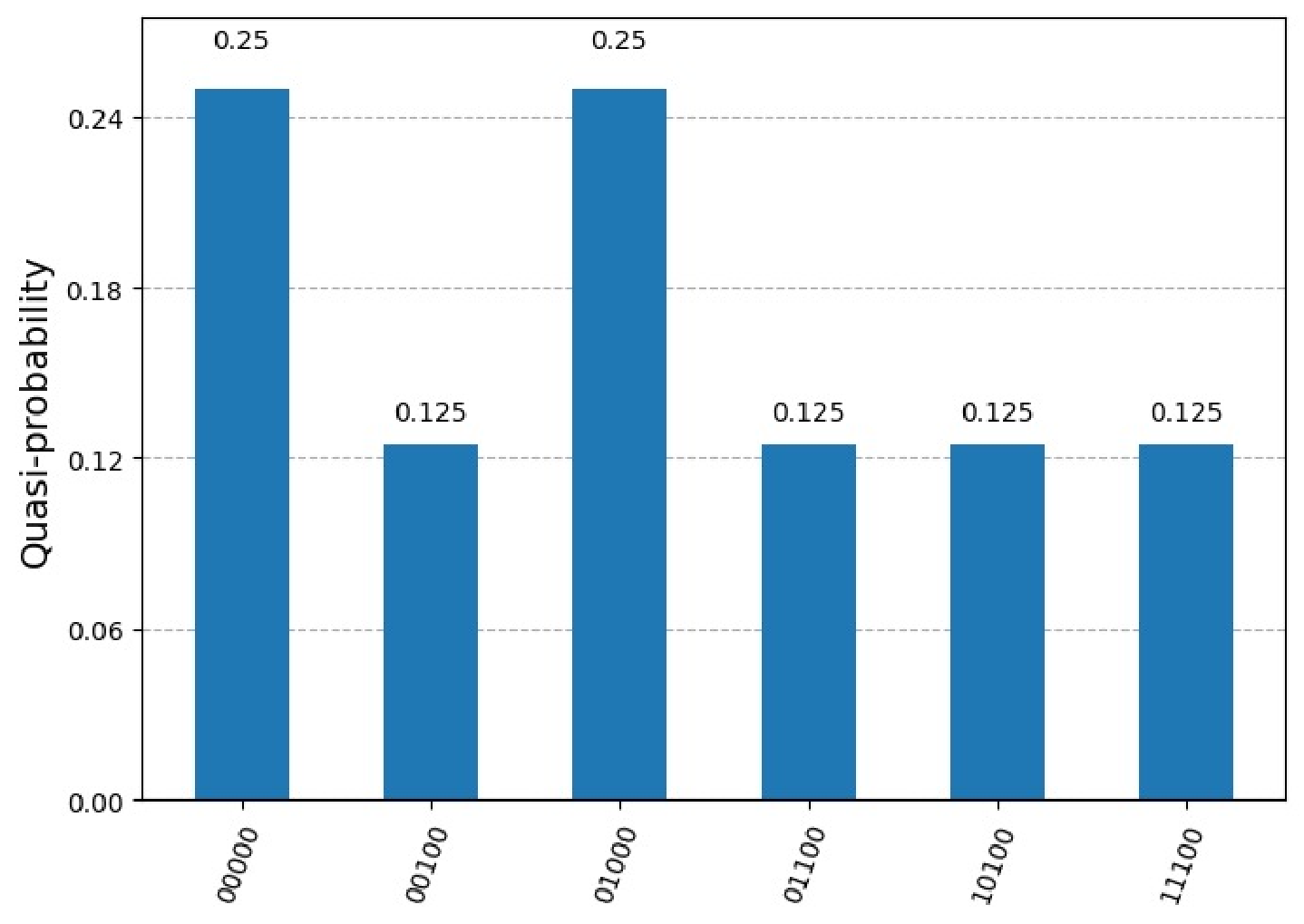}}
%\end{center}
\caption{(a) Quantum circuit with $m=2$ and with $Pred(1)=2$, $Pred(2)=4$, $Pred(3)=2$ with disentanglement. (b) The resulting distribution, the qubits $0$ and $1$ are un-computed to the value $0$. They represent thesquare of the inverse amplitude of the leaves. }
\label{prod_1.eps}
\end{figure}

For a search tree represented in Figure \ref{tree_m2.eps} we generate the superposition by controlled Hadamard gates $cH$ by $ \hat b_i $ with $Pred(1)=2$,  $Pred(2)=4$,  $Pred(3)=2$ determining $| \hat b_i \rangle$ for the node $v$ by 
\[  |\psi_1 \rangle = |1 \rangle | 0 \rangle =  | 1 \rangle Pred(1) | 0 \rangle = | 1 \rangle |  2 \rangle =cH | 1 \rangle |  2 \rangle, 
\]
\[  |\psi_2 \rangle = |2 \rangle | 0 \rangle =  | 2 \rangle Pred(2) | 0 \rangle = | 2 \rangle |  4 \rangle =cH^{\otimes  2} | 2 \rangle |  4 \rangle,  \]
\[  |\psi_3 \rangle = |3 \rangle | 0 \rangle =  | 3 \rangle Pred(3) | 0 \rangle = | 3 \rangle |  2 \rangle =cH | 3 \rangle |  2 \rangle.
\]
In our circuit represented in Figure \ref{prod_1.eps} (a) $Pred(v)$ the values are represented by the unary coding in the qubits $0$  and $1$.
We disentangle $|v\rangle|\hat{b}_i\rangle$ by recomputing with $Pred(v)^{-1}$. The resulting distribution of amplitudes is no longer uniform, see Figure \ref{prod_1.eps} (b).

\subsection{Binomial Distribution of 8-Puzzle} \label{ap: Binomial Distribution}

For the leaf amplitude values, precise path descriptors are not required. We begin with the desired state $v_{desired}$ and explore all possible paths for search depth $m$. The resulting description follows a binomial distribution. 
We assume that the empty space is at an edge. At an edge the path splits evenly in  two possible routes,   each described of four paths:
\begin{itemize}
\item 
Route via corner
\[
edge  \rightarrow  corner \rightarrow  edge
\]
with two 2 available corners, and each corner splits into $(b_3=2$) 2 moves. Total paths created  is $2 \cdot 2 = 4$.
\item Route via center 
\[edge  \rightarrow center \rightarrow edge\]
one center, which splits into 4 ($b_1=4$) moves. Total paths created  is $1 \cdot 4 = 4$.
\end{itemize}
Akin to a fair coin akin that can be described by binomial distribution. The probability of getting exactly $r$ successes in $k$ independent Bernoulli trials for a fair coin with $p=\frac{1}{2}$ is given by
\[
Pr(X,r)=\binom{k}{r} \cdot  \frac{1}{2} ^r \cdot \left(1- \frac{1}{2} \right)^{k-r} =\binom{k}{r} \cdot \frac{1}{2} ^r \cdot \frac{1}{2}^{k-r}.
\]
Since both  paths create a route of four paths after four steps $4^k$ paths are created with 
\[
4^k \cdot \binom{k}{r} \cdot \frac{1}{2} ^r \cdot \frac{1}{2}^{k-r}.
\]
Our initial step starts at the corner which includes  one move  to an edge with $b_1=2$, and creates 2 paths resulting in indicating the number of generated leaves (frequency),
\[
F_r=2 \cdot 4^k \cdot \binom{k}{r} \cdot \frac{1}{2} ^r \cdot \frac{1}{2}^{k-r}=2^{k+1}  \cdot \binom{k}{r} 
\]
with
\[
r=0,1,2,\dots,k.
\]

For depth $m$, we assume $m = 2 \cdot k+2 $ where  $k$ = number of complete 2-step routes. However, $m$ does not need to be an even number ad  we can rewrite it as
$m-1 = 2k+s$ where
$s\in\{0,1\}$    indicates is there is one extra move left over and we can approximate $k$ in relation to $m$ with
\[
k=\left\lfloor \frac{m-1}{2}\right\rfloor.
\]
indicating the correct frequencies $F_r$.

\subsubsection{Costs}

The resulting costs for deterministic padding
\begin{itemize}
\item 
Route via corner: The path moves
\[
edge  \rightarrow  corner \rightarrow  edge
\]
The edge multiplies with $\hat b_2=4$ ($b_2=3$) and the corner with $\hat b_3=b_3=2$.
\item Route via center: The path moves 
\[
edge  \rightarrow center \rightarrow edge
\]
The edge multiplies with $\hat b_2=4$ ($b_2=3$) and the center multiplies with $\hat b_1=b_1=4$. 
\end{itemize}
The total cost  $\hat \Omega_r $ is defined where $r$ is the number of times the route via the center is chosen and $k-r$
is the number of times the route via the corner is chosen, multiplied over two paths resulting in
 \[
\hat \Omega_r = \left( \hat b_2  \cdot \hat b_2 \right) ^{r}\cdot  2 \cdot  \left( \hat b_1 \cdot  \hat b_2 \right) ^{k-r}  =  16^r  \cdot 2 \cdot 8^{k-r}
 \]
 We can convert the values $\hat \Omega_r $ in to  frequency approach used by dynamic pumping  with resulting number of repetitions $\hat \kappa_r$
\begin{equation}
\hat \kappa_r=\frac{\left(\hat b_{max}\right)^k}  { \hat \Omega_r } =\frac{4^k}  {\hat \Omega_r }.
\end{equation}
For probabilistic padding
\[
 \Omega_r = \left(  b_2 \cdot b_3 \right) ^{r}\cdot  2 \cdot  \left(  b_2  \cdot b_1 \right) ^{k-r}  =  12^r  \cdot 2 \cdot 6^{k-r}.
 \]
with
\begin{equation}
\kappa_r=\frac{\left(\hat b_{max}\right)^k}  { \Omega_r } =\frac{4^k}  {\Omega_r }.
\end{equation}

\subsection{Impact of Unique Solutions - 8-puzzle $m=5, 10, 15$} \label{ap: DFS}

Table  \ref{Binominal_5_DFS_Det},  \ref{Binominal_10_DFS_Det} and \ref{Binominal_15_DFS_Det} reports the  values of  $\hat \Omega$  estimated by a binomial distribution, alongside the unique solutions found by DFS  for deterministic state approach. 

\begin{table} [h!]
\caption{For an 8-puzzle search tree of depth  $m=5$,  $\hat \Omega$  is estimated by a binomial distribution, alongside the unique solutions found by DFS with their corresponding appearance probabilities $Pr$.}
\begin{center}
\begin{tabular} { |c | c || c | c | }\hline
 \multicolumn{2}{|c||} {Binomial}   &  \multicolumn{2}{|c|}  {Unique Solutions}  \\ \hline
 $\hat \Omega$  &   $Pr$   &  $\hat \Omega$  &   $Pr$ \\ \hline \hline
- & - & 13 & 0.0.0667  \\ \hline
 - & - & 21 &  0.0.0667  \\ \hline
- & - & \textbf{128}& 0.0.0667  \\ \hline
\textbf{128} & 0.25  & 37 & 0.2 \\ \hline
\textbf{ 256} & 0.5 &  \textbf{256} & 0.6 \\ \hline
 512 & 0.25 & - & - \\ \hline
\end{tabular}
\end{center}
\label{Binominal_5_DFS_Det}
\end{table}
\begin{table} [h!]
\caption{For an 8-puzzle search tree of depth  $m=15$,  $\hat \Omega$  is estimated by a binomial distribution, alongside the unique solutions found by DFS with their corresponding appearance probabilities $Pr$. }
\begin{center}
\begin{tabular} { |c | c || c | c | }\hline
 \multicolumn{2}{|c||} {Binomial}   &  \multicolumn{2}{|c|}  {Unique Solutions}  \\ \hline
 $\hat \Omega$  &   $Pr$   &  $\hat \Omega$  &   $Pr$ \\ \hline \hline
 - & - & $\cdots$ & $\cdots$ \\ \hline
 4194304  &  0.0078 & $\cdots$ & $\cdots$ \\ \hline
 8388608 &  0.05467  & 818401  & 0.0604\\ \hline
 \textbf{16777216} &  0.1641& \textbf{16777216}  & 0.1700\\ \hline
\textbf{33554432}  & 0.2734 &\textbf{33554432} & 0.3017\\ \hline
 67108864 &  0.2734 & - & -  \\ \hline
134217728 &  0.1641 & - & -\\ \hline
 268435456&   0.05467 & - & -\\ \hline
 536870912 &  0.0078  & - & - \\ \hline
\end{tabular}
\end{center}
\label{Binominal_15_DFS_Det}
\end{table}
Table  \ref{Binominal_5_DFS_Prob} ,  \ref{Binominal_10_DFS_Prob} and \ref{Binominal_15_DFS_Prob}  reports the  values of  $ \Omega$  estimated by a binomial distribution, alongside the unique solutions found by DFS  for probabilistic state approach. 
\begin{table} [h!]
\caption{For an 8-puzzle search tree of depth  $m=5$,  $\Omega$  is estimated by a binomial distribution, alongside the unique solutions found by DFS with their corresponding appearance probabilities $Pr$.}
\begin{center}
\begin{tabular} { |c | c || c | c | }\hline
 \multicolumn{2}{|c||} {Binomial}   &  \multicolumn{2}{|c|}  {Unique Solutions}  \\ \hline
 $ \Omega$  &   $Pr$   &  $ \Omega$  &   $Pr$ \\ \hline \hline
- & - & 4 & 0.0.0667  \\ \hline
 - & - & 12 &  0.0.0667  \\ \hline
- & - & \textbf{72}& 0.0.0667  \\ \hline
\textbf{72} & 0.25  & 21 & 0.2 \\ \hline
\textbf{ 144} & 0.5 &  \textbf{144} & 0.6 \\ \hline
 288 & 0.25 & - & - \\ \hline
\end{tabular}
\end{center}
\label{Binominal_5_DFS_Prob}
\end{table}
\begin{table} [h!]
\caption{For an 8-puzzle search tree of depth  $m=10$,  $\Omega$  is estimated by a binomial distribution, alongside the unique solutions found by DFS with their corresponding appearance probabilities $Pr$.}
\begin{center}
\begin{tabular} { |c | c || c | c | }\hline
 \multicolumn{2}{|c||} {Binomial}   &  \multicolumn{2}{|c|}  {Unique Solutions}  \\ \hline
 $\Omega$  &   $Pr$   &  $ \Omega$  &   $Pr$ \\ \hline \hline
   - & -  & $\cdots$ & $\cdots$ \\ \hline
7776 &  0.0625  &  1197 & 0.0649\\ \hline
\textbf{15552}  & 0.25 &  \textbf{15552}  &  0.2035 \\ \hline
 \textbf{31104} & 0.375 & \textbf{31104}  & 0.3896 \\ \hline
62208 & 0.25 & - & -  \\ \hline
 124416 &  0.0625 & - & - \\ \hline 
\end{tabular}
\end{center}
\label{Binominal_10_DFS_Prob}
\end{table}
\begin{table} [h!]
\caption{For an 8-puzzle search tree of depth  $m=15$,  $\Omega$  is estimated by a binomial distribution, alongside the unique solutions found by DFS with their corresponding appearance probabilities $Pr$.}
\begin{center}
\begin{tabular} { |c | c || c | c | }\hline
 \multicolumn{2}{|c||} {Binomial}   &  \multicolumn{2}{|c|}  {Unique Solutions}  \\ \hline
 $\Omega$  &   $Pr$   &  $ \Omega$  &   $Pr$ \\ \hline \hline
 - & - & $\cdots$ & $\cdots$ \\ \hline
 559872 &  0.0078 & $\cdots$ & $\cdots$ \\ \hline
 1119744 &  0.05467 & 109243  & 0.0604\\ \hline
 \textbf{2239488} & 0.164 & \textbf{2239488} & 0.1700\\ \hline
\textbf{4478976} & 0.2734  & \textbf{4478976} & 0.3017\\ \hline
8957952 &  0.2734 & - & - \\ \hline
 17915904 & 0.1641 & - & - \\ \hline
 35831808 &  0.05467 & - & -  \\ \hline
71663616	 &  0.0078 & - & - \\ \hline
\end{tabular}
\end{center}
\label{Binominal_15_DFS_Prob}
\end{table}

\subsection{Nested Search with Amplitude Amplification}  \label{ap: NEA}

The idea of saving the costs using heuristics and amplitude amplification in quantum tree search is based on the inequality, that represents the  decomposition  of the  Hilbert space $\mathcal{H}$ of dimension $n=2^m$ into
subspaces  $\mathcal{L}$ and  $\mathcal{U}$ of dimension $2^g$ and $2^{m-g}$. 
The costs of Grover's algorithm for the quantum tree search algorithm are
\[\sqrt{2^m}=\sqrt{2^g} \cdot \sqrt{2^{m-g}}.\]
Upon initial inspection, it appears that amplitude amplification could potentially reduce costs significantly to
 \begin{equation}
\sqrt{2^g} + \sqrt{2^{m-g}}. 
 \end{equation}
In contrast, the cost associated with Grover’s algorithm for the quantum tree in the subspace $\mathcal{L}$ would be $\sqrt{2^g/k}$, resulting in $k$ solutions. Consequently, the resulting cost of the amplitude amplification for the subspace $\mathcal{U}$ would be $\sqrt{2^m}$, as follows:
\[
\sqrt{2^g/k} + 2 \cdot cost(A) \cdot \sqrt{k \cdot 2^{m-g}} = \sqrt{2^g/k} + 2 \cdot \sqrt{2^g/k} \cdot \sqrt{k \cdot 2^{m-g}} =
 \]
 \begin{equation}
 =  \sqrt{2^g/k} +  \sqrt{2^g/k} \cdot \sqrt{2^{m}} =  \sqrt{2^g/k}  \cdot (1 + \sqrt{2^{m}} )
 \end{equation}
 This results in even higher costs by a factor of $\sqrt{2^g/k}$ with
\begin{equation}
\sqrt{2^g/k} \cdot (1 + \sqrt{2^{m}}) > \sqrt{2^{m}}
\end{equation}
since we cannot ignore the preparation cost $A \cdot | 0^{\otimes m} \rangle$.

\bibliographystyle{apalike}

\end{document}